# Structural, electric and magnetic properties of the electron-doped manganese oxide: $La_{1-X}Te_XMnO_3$ (x=0.1, 0.15)

G.T.Tan, S.Y.Dai, P.Duan, Y.L.Zhou, H.B.Lu, Z.H.Chen[*]

Laboratory of Optical Physics, Institute of Physics & Center for Condensed Matter Physics, Chinese Academy of Sciences, P.O. BOX 603, Beijing 100080, P. R. China

**Abstract**

In this study, the electrical and magnetic properties of $La_{1-x}Te_xMnO_{3-\delta}$ (x=0.1, 0.15), which is a new material and shows good colossal magnetoresistance (CMR) behavior, have been investigated. These compounds have rhombohedral structure. In this material, Te replaced a part of La ions, which induced the lattice cell constriction and Mn-O-Mn bond angle widening. X-ray photoemission spectroscopy (XPS) measurement revealed that the Te ions were in the tetravalent state and the manganese ions could be considered as in a mixture state of $Mn^{2+}$ and $Mn^{3+}$. Thus the material could be viewed as an electron-doped compound. The Curie temperature (Tc) was about 240 K and 255 K for x=0.1, 0.15, respectively. The maximum magnetoresistance ratio MR= $[\rho(0) - \rho(H)]/\rho(0)$ was about 51% at 200 K and in the applied magnetic field of 40 kOe.

[*] Correspondence should be addressd to: Prof. Z.H.Chen, Laboratory of Optical Physics, Institute of Physics & Center for Condensed Matter Physics, Chinese Academy of Sciences, P. O. BOX 603, Beijing 100080, P. R. China.
Email address: zhchen@aphy.iphy.ac.cn  FAX: 86-10-82649531

## I. INTRODUCTION

Since the end of last century, the colossal magnetoresistance (CMR) phenomenon in Perovskite manganites has become an important research subject for its potential applications.[1] The resistance of this kind of compound is a sensitive function of magnetic field. The metal-insulator (MI) transition and related magnetic ordering transition taken place at low temperature exist also in this compound. Traditionally, the interpretation of the CMR phenomenon was based on the double-exchange (DE) model[2] and Jahn-Teller distortion effects.[3] Generally, the manganese ions in $LaMnO_3$ family compound are considered as in a mixed-valence state of $Mn^{3+}$-$Mn^{4+}$. The $Mn^{3+}$ ion has the electron configuration $t_{2g}^3 e_g^1$. The low energy $t^3_{2g}$ triplet state contributes a local spin of S=3/2, while $e_g$ electron is either itinerant or localized depending on the local spin orientation. The $e^1_g$ electron results in an orbital degenerate state and is Jahn-Teller active. As a magnetic field applied, the $e_g$ itinerant electrons hop between $Mn^{3+}$ and $Mn^{4+}$, and the hopping probability depends on the Mn-O-Mn bond angle and the relative orientation of the local spin. This kind of physical picture qualitatively depicted the CMR magneto-resistance behavior. Most of Perovskite manganites can be described as $R_{1-x}A_xMnO_3$, where R is rare earth element, A is divalent metal and x is the doping level. These manganites, which have been widely studied, display excellent CMR behavior at low temperature in strong magnetic field. However, they usually are hole-doped compounds and show a mixed-valence state of $Mn^{3+}$-$Mn^{4+}$. Recently, S.Das[4,5] and J.R.Gebhard[6] reported the CMR behavior in the $La_{1-x}Ce_xMnO_3$ system that was electron-doped compounds. Their studies suggested that the CMR behavior probably occurred in a system of a mixed-valence state of $Mn^{2+}$-$Mn^{3+}$. In the $La_{1-x}Ce_xMnO_3$ material, Ce is a rare-earth ion with tetra-valence. Therefore, it is of great interest to investigate whether the CMR effect exists in the compounds in which $La^{3+}$ is partially replaced by a tetra-valence element in Chalcogen rather than rare-earth element. In this work, we made efforts to study the substitution effect of Tellurium for a part of La in Perovskite manganites.

## II. EXPERIMENT

Polycrystalline samples of $La_{1-x}Te_xMnO_3$ (x=0.1, 0.15) were prepared by conventional ceramic techniques. The stoichiometric mixture of high pure $La_2O_3$, $TeO_2$, and $Mn_3O_4$ powders was milled, pressed and pre-heated at 700 $^0$C for 24 hours, and then the sample was ground and fired at 900 $^0$C for 12 hours. Both the steps mentioned above were processed in the flowing argon gas. In the end the sample was sintered at about 930 $^0$C for 24h in the flowing oxygen gas and then cooled down to room temperature in the off-powered furnace. X-ray diffraction (XRD) measurement was performed with DMAX2400 diffractometer and Cu $K\alpha$ radiation. The oxygen content was analyzed by the Oxygen/Nitrogen Determinator of LECO TC-300. The magnetization measurement was carried out with a superconducting quantum interference device (SQUID) magnetometer and the electrical resistance was measured using a standard four-probe method in the temperature range of 5-300 K. The X-ray photoemission spectrum (XPS) analysis of $La_{0.9}Te_{0.1}MnO_3$ was performed using Al-$K\alpha$ radiation on the EscaLab 220-IXL electronic spectrometer and the scanning step is 0.05 eV.

## III. RESULTS AND DISCUSSION

### A. Oxygen content and structure analyses

If the δ represented the deviation of the oxygen content in our samples, its values determined by the LECO TC-300 Oxygen/Nitrogen Determinator were +0.01 and +0.02 for x=0.1 and 0.15, respectively. The rich oxygen was possible due to the samples being sintered in the flowing oxygen. Compared with the doped level, the effect of the δ was not principle for our samples. We neglected the δ in the chemical formulas but to stand out Te-doped effect.

The result of XPS measurement (shown in Fig.1) revealed that the binding energies of the Te-$3d_{5/2}$ and Te-$3d_{3/2}$ electrons are about 575.9 eV and 586.4 eV, respectively. The value is very close to the standard value of Te-3d in $TeO_2$. Thus the Tellurium ions in this compound were in the tetravalent state and the manganese ions could be considered as in an $Mn^{3+}$ and $Mn^{2+}$ mixture state. The $La_{1-x}Te_xMnO_3$ compounds could be n-type materials and the conductivity could be an electronic one.

It is known that the radius of the $Te^{4+}$ is about 0.089 nm and that of the $La^{3+}$ is 0.1016 nm. The part substitution of $Te^{4+}$ ions for $La^{3+}$ ions will cause the cell constriction and the distortion of lattice, which has been confirmed by the XRD pattern. As shown in Fig. 2, the compound has rhombohedral structure with the space group $R\bar{3}c$. The structure parameters were refined using the DBW9411 program and obtained as a=0.5520 nm for x=0.1 and a=0.5513 nm for x=0.15. The Mn-O-Mn bond angle, that were determined by the diamond program, was about 165.37° and 165.69° for x=0.1 and 0.15 samples, respectively. These structure parameters suggested that the lattice constriction and the bond angle increased slightly with increasing x.

**B. Electric and magnetic properties**

The curves of the dM/dT vs T for $La_{1-x}Te_xMnO_3$ under the applied magnetic field of 10 kOe are shown in Fig. 3a. It shows that the para- to ferro-magnetic transition temperatures Tc are about 240 K for x=0.1 and 255 K for x=0.15, respectively. The increase of Tc with increasing Te concentration are maybe able to attribute to the strengthening of the interactions of both the lattice constriction and bond angle increase. These interactions strengthened the ferromagnetic coupling between Mn ions. The temperature dependences of the resistivity for the samples in the temperature range of 5 to 300 K are plotted in Fig. 3 (b). The metal-semiconductor (MS) transition temperature Tp, where the resistivity ρ is maximum, is about 200 K for x=0.1, and 220 K for x=0.15 in the external magnetic field of 40 kOe. The value was lower than that of the corresponding Tc. On the other hand, Tp shifted to higher temperature and the resistivity reduced significantly when a magnetic field was applied to the sample, or the Te ions content increased. This could be considered as a result of the increase of the carrier density with increasing Te concentration. The variation of doped level also changed the bond angle and widened the single electron $e_g^1$ bandwidth that strengthened the double exchange interaction. Besides, the effect of magnetic field was due to spin moment ordering in external magnetic field, and gave rise to the magnetoresistance effect in the compounds. The magnetoresistance ratio (MR) was defined as $\Delta\rho/\rho(0) = [\rho(0) - \rho(H)]/\rho(0)$, where $\rho(0)$ and $\rho(H)$ is the resistivity of zero field and field H, which also plot in the inset of Fig. 3b. The MR of $La_{0.9}Te_{0.1}MnO_3$ in 40 kOe is about 51%, and $La_{0.85}Te_{0.15}MnO_3$ about 47%.

Fig. 4a shows the magnetic field dependence of the magnetization at 5 K for $La_{1-x}Te_xMnO_3$. The curves sharply increase at the beginning and then approach a saturation value with magnetic field increasing. This phenomenon traditionally was considered as a result of the rotation of the magnetic domain under the action of magnetic field. The saturation moment Ms, as shown in Fig.4 (a), are obtained about 60 emu/g for x=0.1 and 50 emu/g for x=0.15. The lower Ms corresponds to the higher doping level. This phenomenon is similar to that observed in another electronic doping compound $La_{1-x}Ce_xMnO_3$,[6] and $La_{1-x}Zr_xMnO_3$,[7] and contributed to the competition between the DE and the core spin interaction, as well as the canting of the core spin moment with the doped level increases. Furthermore, the magnetic field (H) dependences of the resistivity ρ are shown in Fig. 3(b). For the sample of x=0.1, figure 4b shows that the resistivity sharply reduces at the interval of low magnetic field that just corresponds to a significant increase of the magnetization. (shown in Fig. 4a) This phenomenon is similar to that observed in the previous reports.[8,9] The reduction of the resistivity was attributed to the rearrangement of the nonaligned domains in a strong external field. As a result, spin tended to align along the field direction and led electrons to across easily the domain-wall boundaries. Hwang et al[9] pointed out that the large MR at low field region originated from the spin-polarized electron tunneling between grains, and at high field the MR was remarkably temperature independence. However, the magnetic field dependence of the resistivity of the $La_{0.9}Te_{0.1}MnO_3$ at high field was dependent on the temperature. The resistivity was a linear function of the magnetic field, but there was different slope for the various temperatures, such as for temperature 5 K and 225 K. Based on the double exchange theory, the compound was considered as in mixed-valence state of Mn ions. The possibility of the $e^1_g$ electron hopping between Mn ions, and the one-electron bandwidth of the $e^1_g$ state depended on the length and angle of the Mn-O-Mn bond which were the function of temperature, pressure, magnetic field and magnetization.[10,11] As a magnetic field applied to the sample, which induced the variations of local spin orientation and angle of the Mn-O-Mn bond, thus the resistivity decreased with the field increasing, and the curve of ρ vs H showed different slope at different temperature

## IV. CONCLUSION

In summary, Te-doped Perovskite manganese oxide is a new colossal Magnetoresistance material. In this compound, we used Tellurium to replace a part of La ions. XPS analysis indicated that the Te ions were in the tetra-valence state. It suggested that the material was an electron-doped CMR compound, and could be in a mixture state of $Mn^{3+}$ and $Mn^{2+}$. The electronic and magnetic properties of the samples were measured, which showed that the increase in Te doped-level drove the Tc from 240 K rising to 255 K, meanwhile the saturation magnetization decrease, as well as the electrical resistivity decrease due to the increases of the carrier density. Besides, a large magnetoresistance at lower magnetic field for $La_{0.9}Te_{0.1}MnO_3$ was observed.

## ACKNOWLEDGMENT

Authors greatly appreciate Dr. LinTao Yang for his help in the structure analysis. This work was supported by a grant for State Key Program No. G1998061412 of China.

# Figure captions

Fig.1. Te-3d core level XPS spectrum of $La_{0.9}Te_{0.1}MnO_3$.

Fig.2. The XRD patterns of $La_{1-x}Te_xMnO_3$ (x=0.10 and 0.15), Space group $R\bar{3}c$

Fig.3 (a) dM/ dT vs T curves of $La_{1-x}Te_xMnO_3$ at 10 kOe.
(b) Resistivity vs temperature curves of $La_{1-x}Te_xMnO_3$ in the magnetic field of 40 kOe. Inset: the corresponding temperature dependences of MR% values.

Fig.4. (a) Field dependences of magnetization of $La_{1-x}Te_xMnO_3$ at 5 K ;
(b) Resistivity versus magnetization of the samples at 5 K, 225 K and 240 K

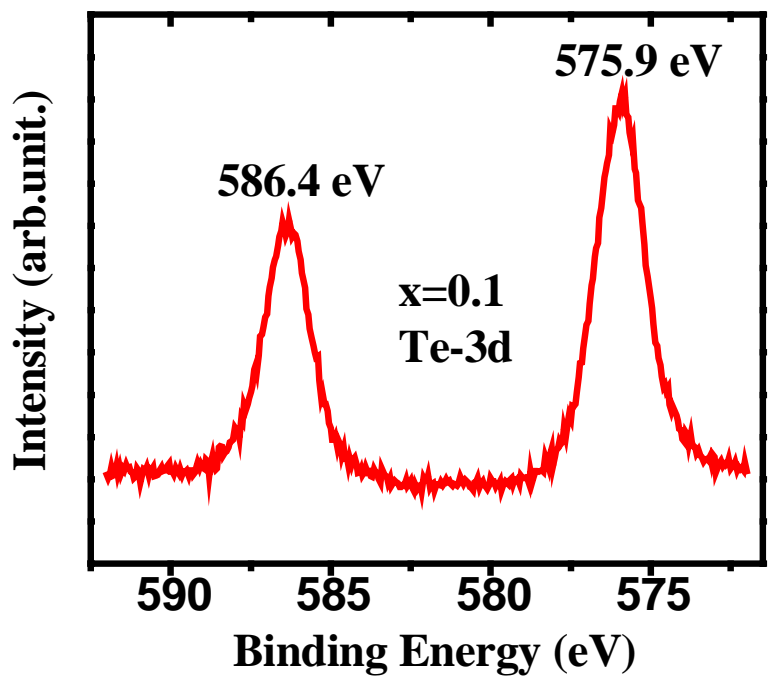

Fig.1. Tan et al submitted to *J.Appl.Phys..*

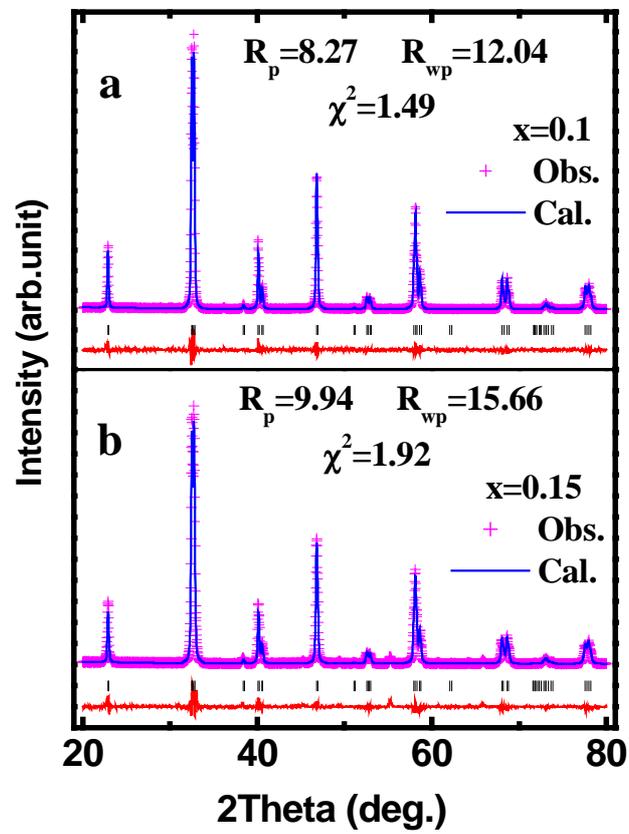

Fig.2. Tan et al submitted to *J.Appl.Phys.*.

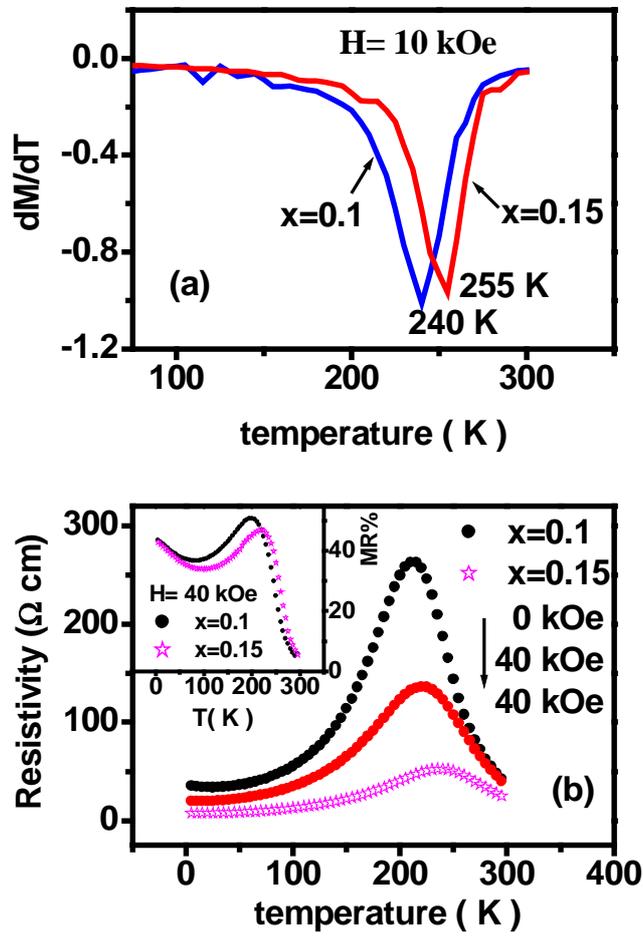

**Fig.3. Tan et al submitted to *J.Appl.Phys.***

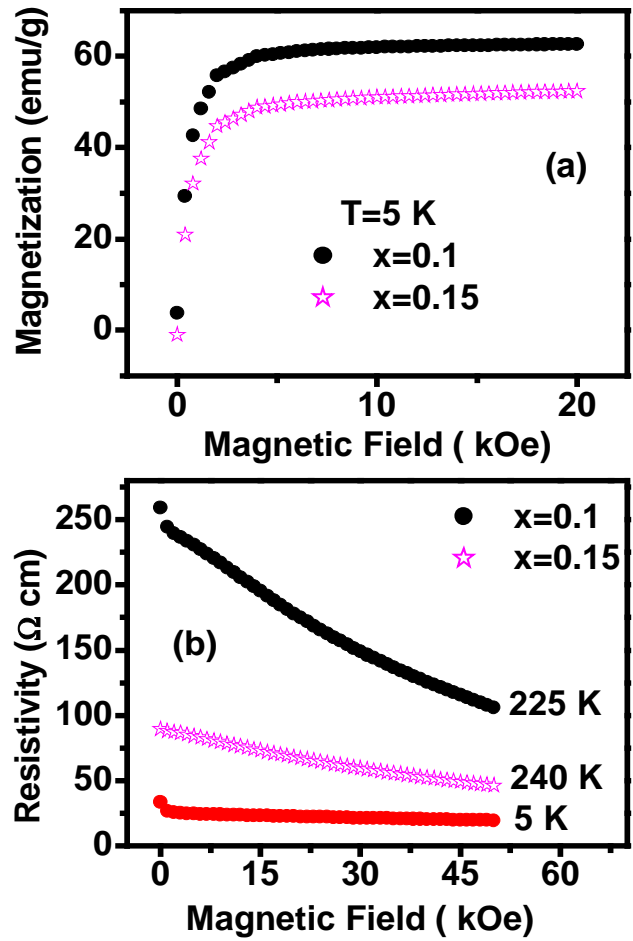

**Fig.4.** Tan et al submitted to *J.Appl.Phys.*